\documentclass[aps,prb,showpacs,twocolumn,amsmath,amssymb,superscriptaddress]{revtex4}
\usepackage{graphicx}
\usepackage{Jonasmacros}
\usepackage{bm}
\usepackage{graphicx}
\usepackage[german,english]{babel}
\usepackage{fancyheadings}
\newcommand{\br}{{\bf r}}

\newcommand{\beqa}{\begin{eqnarray}}
\newcommand{\eeqa}{\end{eqnarray}}

\begin{document}
\date{\today}
\title{Surface imaging of inelastic Friedel oscillations}
\author{J. Fransson}
\email{jonasf@lanl.gov}
\affiliation{Theoretical Division, Los Alamos National Laboratory, Los Alamos, New Mexico 87545, USA}
\affiliation{Center for Nonlinear Studies, Los Alamos National Laboratory, Los Alamos, New Mexico 87545, USA}
\author{A. V. Balatsky}
\email{avb@lanl.gov}
\affiliation{Theoretical Division, Los Alamos National Laboratory, Los Alamos, New Mexico 87545, USA}
\affiliation{Center for Integrated Nanotechnology, Los Alamos National Laboratory, Los Alamos, New Mexico 87545, USA}

\begin{abstract}
Impurities that are present on the surface of a metal often have
internal degrees of freedom.  Inelastic scattering   due to
impurities   can be revealed by  observing local  features seen in
the tunneling current with
   scanning
 tunneling microscope (STM). We consider
 localized vibrational modes coupled to the electronic
  structure of a surface. We argue that
  vibrational modes of impurities produce  Fermi momentum $k_F$
  oscillations in second derivative of
  current with respect to voltage $\partial^2I(\br, V)/\partial V^2$. These
  oscillations are similar to the well known Friedel
  oscillations of screening charge on the surface.
   We propose to measure inelastic scattering  generated by the
    presence of the vibrational modes
   with   STM by imaging the $\partial^2I/\partial V^2$
   oscillations on the metal surface.
\end{abstract}
 \maketitle

\section{Introduction}
\label{sec-introduction}

Inelastic electron tunneling spectroscopy (IETS)
with scanning tunneling microscope, IETS-STM, is  a well-established
technique, starting with an important experiment by Stipe
\etal\cite{stipe1998}
 These experiments show step like features in the tunneling
 current and local density of states (DOS). The physical
 explanation of the effect is straightforward: once the energy
 of the tunneling electron exceeds the energy required to
 excite local vibrations, there is a new scattering process
 that contributes to the scattering of electrons due to
  inelastic excitation of the local mode.\cite{jaklevich1966}
   The local vibrational modes can be seen in differential
    conductance measurements as side-bands to the main
    (elastic) conductance peak.\cite{park2000} Recently,
    IETS-STM was
    used to measure the spin excitations of individual magnetic
     atoms.\cite{heinrich2004} In addition, recent experiments
      show dominant inelastic channels which are strongly
      spatially localized to particular regions of a
      molecule.\cite{grobis2005}

Improvements in spectroscopic and microscopic measurements have provided new information about fundamental aspects of scattering and interactions in the solid state. Recently, STM technique has evolved from imaging of surfaces,\cite{binnig1987} to spin polarized tips for magnetic sensitivity of the read-out\cite{heinrich2004} and spin-polarized injection of the tunneling current used in the measurements.\cite{wiesendanger} STM has been also  used to detect Kondo interactions between conduction electrons and single atomic spins\cite{madhavan1998,li1998} and to study the properties of individual atoms.\cite{yamasaki2003,bode2004}

We propose to use  resolution of STM to address spatially resolved
inelastic tunneling features produced by local inelastic scattering
of the molecule  both near the scattering center and at far
distances. The "fingerprint" of the inelastic scattering will be
present even away from the scattering center. Using local IETS-STM
would enable measurements of the standing waves produced by
    inelastic scattering off impurities,
    which could be revealed as waves  seen in the {\em second derivative}
    $\partial^2I(\br, V)/\partial V^2$. For the simplest model of parabolic
    conduction band these waves will be seen as standing waves with
    period set by the Fermi momentum $k_F$.  These standing waves are seen
    in the oscillations of the $\partial^2I(\br, V)/\partial V^2$ and are
    {\em similar but qualitatively different} from the conventional
    Friedel oscillations.

     In case of regular Friedel oscillations
      the charge that screens off the impurity exhibit
 oscillations at large distances from the impurity.
 Recent
  STM measurements have observed Friedel oscillations in the
  elastic scattering channels on surfaces with atomic
  impurities adsorbed on the
  surface.\cite{madhavan1998,morgenstern2004} These oscillations are
  seen in a wide range of bias as they reflect screening of the
  charge by electron states in the whole bandwidth.
Inelastic scattering of surface electrons off the
    molecules   may be
   viewed as {\em inelastic} Friedel oscillations produced by the
   electron states that are involved in screening.
  Inelastic Friedel oscillations are seen only in the narrow window of
  energies near the energy of the mode $\omega_0$ at which inelastic
  scattering occurs, in contrast to conventional Friedel screening.

It was recently proposed to use IETS-STM to address the inelastic scattering features in the Bi2212 superconductors.\cite{zhu2006,balatskyRMP2006} In the case of superconductors, the situation is more involved since one has to deal with the more complicated band structure of Bi2212 and with the inelastic scattering off the distributed bosonic modes. The work presented here allows one to test similar ideas in a more controlled set up where the metal surface and molecular modes are well understood.

The paper is organized as follows.
In Sec. \ref{sec-model} we introduce
a model which describe the addressed
issue and some numerical results are discussed in
Sec. \ref{sec-results}. The paper is concluded in Sec. \ref{sec-conclusions}.

\section{Probing the Friedel oscillations and inelastic scattering measurements}
\label{sec-model} The system we consider consists of a
 two-dimensional surface on which
 inelastic scattering centers are
  randomly distributed. For simplicity,
   we assume that the impurities are distributed sufficiently
    far from each other so that their mutual interactions can
     be neglected. All vibrational modes have energy $\omega_0$ and are
      assumed to be the same as they would come from the same type of molecules.
     We use the Hamiltonian for the local vibrational
     modes, coupled to electrons via Holstein coupling with interactions assumed
     to occur only at the single impurity site, so that
\begin{equation}
\Hamil=\sum_{\bfk\sigma}\dote{\bfk}\cdagger{\bfk}\cc{\bfk}
    +\omega_0\bdagger{}\bc{}
    +\lambda\sum_{\bfk\bfk'\sigma}\csdagger{\bfk\sigma}\cs{\bfk'\sigma}(\bdagger{}+\bc{}).
\label{eq-Ham}
\end{equation}
Here, a surface electron is created (annihilated) by $\cdagger{\bfk}\ (\cc{\bfk})$ at the energy $\dote{\bfk}$. The strength of the electron-phonon interaction is given by the parameter $\lambda$, whereas $\omega_0$ is the mode of the bare phonon which is created (annihilated) by $\bdagger{}\ (\bc{})$.

The features we are considering should be seen in the second
derivative of the tunneling current with respect to the bias voltage
$V$ in real space, i.e. $\partial^2I(\bfr,V)/\partial V^2$. This
quantity is directly proportional to the frequency derivative of the
local DOS. In second order perturbation theory,
this amounts to taking the frequency derivative of the correction to
the density of states, $\delta N(\bfr,\omega)$, due to the influence
of the impurity scattering.  The real space electron Green function (GF)
is given by
\begin{equation}
G(\bfr,\bfr';\omega)=G_0(\bfr,\bfr';\omega)
    +G_0(\bfr,0;\omega)\Sigma(\omega)G_0(0,\bfr';\omega),
\label{eq-GF}
\end{equation}
with the zero GF in two spatial dimensions
\begin{eqnarray}
\lefteqn{
G_0(\bfr,\bfr';\omega)=
    \sum_\bfk G_0(\bfk;\omega)e^{i\bfk\cdot(\bfr-\bfr')}
}
\nonumber\\&=&
    2\pi\frac{m}{\hbar^2}J_0(k_F|\bfr-\bfr'|)
    \biggl(\log\biggl|\frac{\omega+D}{\omega-D}\biggr|-i\pi\biggr).
\label{eq-G0}
\end{eqnarray}
Here, the bare electron Fourier transformed GF $G_0(\bfk,\omega)=1/(i\omega-\dote{\bfk})$. Using standard procedure for frequency summation, the self-energy $\Sigma(\omega)$ is found as
\begin{eqnarray}
\lefteqn{
\Sigma(i\omega)=-\frac{\lambda^2}{\beta}\sum_{\bfk,m}
    G_0(\bfk,i\omega+i\Omega_m)D_0(i\Omega_m)
}
\label{eq-S}\\&=&
    \lambda^2\sum_\bfk\biggl\{
        \frac{n_B(\omega_0)+f(\dote{\bfk})}{i\omega+\omega_0-\dote{\bfk}}
        +\frac{n_B(\omega_0)+1-f(\dote{\bfk})}{i\omega-\omega_0-\dote{\bfk}}
        \biggr\},
\nonumber
\end{eqnarray}
where $D_0(i\omega)=2\omega_0/(\omega^2-\omega_0^2)$ is the bare phonon GFs.

Due to these definitions, it is clear that the correction to the local DOS is given by
\begin{equation}
\delta N(\bfr,\omega)=\frac{1}{\pi}\im G_0(\bfr,0,\omega)\Sigma(\omega)G_0(0,\bfr,\omega)
\label{eq-deltaN}
\end{equation}
Since the effects from inelastic scattering are included in the
self-energy, we evaluate $\partial\delta N(\omega)/\partial\omega$,
which is directly proportional to $\partial^2I(\bfr,V)/\partial V^2$.

Sharp feature in  self-energy $\im\Sigma(\omega)$ is connected to
the inelastic scattering feature. Noting that $\re
G_0(\bfr,\bfr',\omega)\approx0$, our calculations simplify to
$\delta
N(\bfr,\omega)\approx(1/\pi)G_0(\bfr,0;\omega)[\im\Sigma(\omega)]G_0(0,\bfr;\omega)$.
We thus find that
\begin{equation}
\delta N(\bfr,\omega)\approx-\pi^3\biggl(\frac{2m}{\hbar^2}\biggr)^2
    J_0^2(k_Fr)\im\Sigma(\omega).
\label{eq-dN}
\end{equation}
Analytical continuation, e.g. $i\omega\rightarrow\omega+i0^+$, of
the self-energy in Eq. (\ref{eq-S}) gives the imaginary part
\begin{eqnarray}
\im\Sigma(\omega)&=&
    -2\pi^2\lambda^2\frac{m}{\hbar^2}[2n_B(\omega_0)+1
\nonumber\\&&
    +f(\omega+\omega_0)-f(\omega-\omega_0)].
\label{eq-imS}
\end{eqnarray}
The correction to the local DOS provides the oscillation in real space. The
electron-phonon interaction gives rise to an increased local DOS for
frequencies $|\omega|>\omega_0$.

We find that $\partial^2I(\bfr,V)/\partial V^2$ is proportional to
\begin{equation}
\frac{\partial}{\partial\omega}\delta N(\bfr,\omega)\approx
    -\pi^3\biggl(\frac{2m}{\hbar^2}\biggr)^2
    J_0^2(k_Fr)\frac{\partial}{\partial\omega}\im\Sigma(\omega),
\label{eq-ddN}
\end{equation}
where
\begin{eqnarray}
\frac{\partial}{\partial\omega}\im\Sigma(\omega)=
    2\pi^2\lambda^2\frac{m}{\hbar}\frac{\beta}{4}\biggl(
        \cosh^{-2}\frac{\beta}{2}(\omega+\omega_0)
\nonumber\\
        -\cosh^{-2}\frac{\beta}{2}(\omega-\omega_0)\biggr).
\label{eq-dS}
\end{eqnarray}
taking
$(\beta/4)\cos^{-2}\beta(\omega\pm\omega_0)/2\rightarrow\delta(\omega\pm\omega_0)$,
as $T\rightarrow0$, we expect to find sharp features in
$\partial^2I(\bfr,V)/\partial V^2$ around $\omega=\pm\omega_0$ for low
temperatures. Hence, our simplified analytical calculations 
show that inelastic impurities that couple to the surface
electrons give rise to Friedel oscillations in the real space image
of $\partial^2I(\bfr,V)/\partial V^2$.

One important point that we  also need  to address is the role of
the dephasing in the inelastic scattering process. We are dealing
with the inelastic scattering process in which the incoming and
outgoing electron waves have different energy. Energy transferred
to/from the inelastic center  implies the phase change  and hence
the interference of the incoming and outgoing waves will be affected
by  dephasing caused by scattering.  This dephasing process would
occur even at $T=0$ and hence has nothing to do with the thermal
scattering. In the limit of small boson energy $\hbar \omega_0 \ll
E_F$ dephasing is small and is not going to destroy the interference
of the incoming and outgoing waves. Qualitatively we can estimate
the change in the phase  as a result of inelastic scattering as
$\delta \phi = \delta \epsilon \delta t$. Here $\delta \epsilon =
\hbar \omega_0$ is the energy transferred to/from the electron and
$\delta t \sim a/v_F$ is the typical time for the collision of
electron with the impurity of size $a$, assumed to be on the unit
cell size. Then we estimate $a = 2\pi/k_F$ and obtain
\begin{equation}
\delta\phi \sim 2\pi \hbar \omega_0/E_F \ll 2\pi.
\label{EQ:phase1}
\end{equation}
Therefore the interference between incoming and outgoing waves will
survive as long as energy transferred in the collision is small.

\section{Results}
\label{sec-results} 
Although the analytical calculations in Sec. \ref{sec-model} clearly shows oscillatory response of the inelastic scattering, we provide in this section some numerical results to emphasize our proposal. The numerical calculations of $\delta N(\bfr,\omega)$ are based on the full expression presented in Appendix \ref{app-dN}.

\begin{figure}[t]
\begin{center}
\includegraphics[width=8cm]{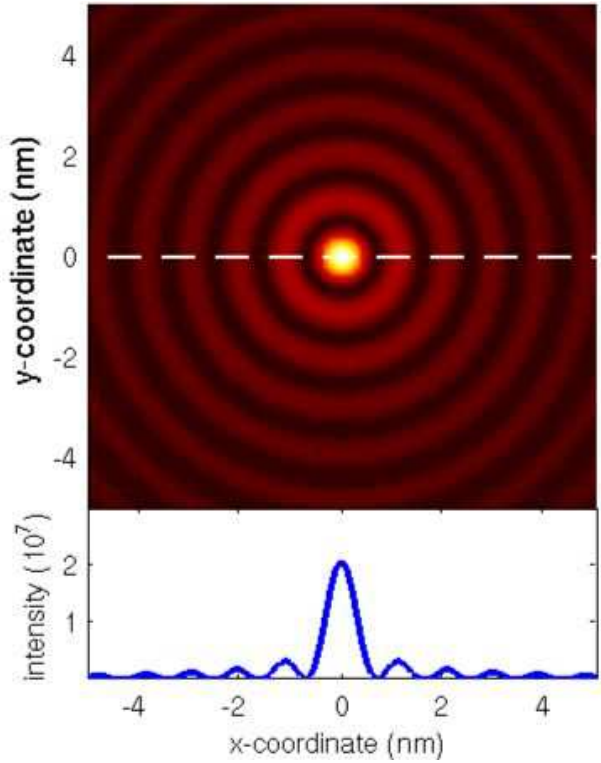}
\end{center}
\caption{(Color online) Spatial dependence of $\delta N(\bfr-\bfr_0,\omega=\omega_0)$, for $T=1$ K.}
\label{fig-onespot}
\end{figure}
\begin{figure}[t]
\begin{center}
\includegraphics[width=8cm]{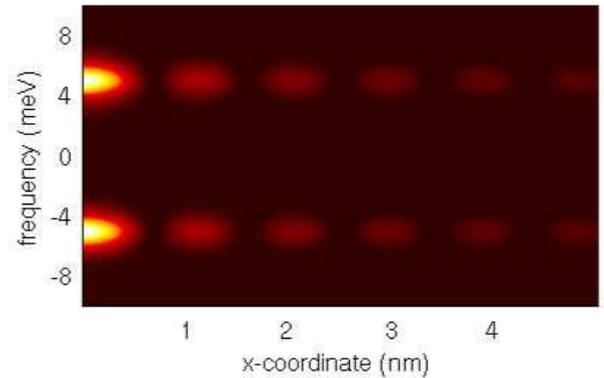}
\end{center}
\caption{Spatial and frequency dependence of $|\partial\delta N(\bfr,\omega)/\partial\omega|$ at $T=5$ K.}
\label{fig-dNR}
\end{figure}
\begin{figure}[b]
\begin{center}
\includegraphics[width=8cm]{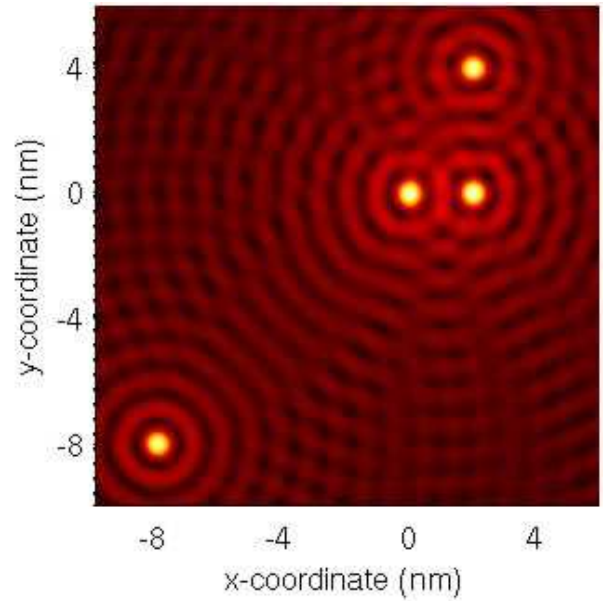}
\end{center}
\caption{(Color online) Calculated IEST-STM resonance, e.g. at phonon mode resonance, signal on surface with four inelastic impurities embedded.} \label{fig-fourspots}
\end{figure}

From Eq. (\ref{eq-ddN}) it is clear that the oscillations only have a radial component, because the assumed electron-phonon coupling is rotationally invariant. This feature is manifested in Fig. \ref{fig-onespot}, where the upper panel shows the spatial dependence of $\delta N(\bfr-\bfr_0,\omega_0)$ at the resonance energy $\omega=\omega_0$. The lower panel shows the intensity of $\delta N(x-x_0,0;\omega_0)$, on the line $\bfr-\bfr_0=x-x_0$, which corresponds to the dashed line in the upper panel. The features in $\delta N(\bfr-\bfr_0,\omega)$  depend on the spatial position \emph{and} on the frequency. In Fig. \ref{fig-dNR}, which shows a contour plot of $\delta N(x-x_0,0;\omega)$, we illustrate the fact that the inelastic Friedel oscillations occur in a narrow interval around $\omega=\pm\omega_0$. The plot clearly illustrates that there is hardly any intensity at all, for frequencies off the phonon mode $\omega_0$.
The plot in Fig. \ref{fig-fourspots} illustrates the IETS-STM resonance signal on a surface with four inelastic scattering centers, analogous to the plot in Fig. \ref{fig-onespot}. As expected from the wave character of the signal, the pattern on the surface is because of interfering Friedel oscillations from the four scattering centers.
%The plot in Fig. \ref{fig-fourspots} shows IEST-STM signal on surface with four inelastic impurities embedded.

\begin{figure}[t]
\begin{center}
\includegraphics[width=8cm]{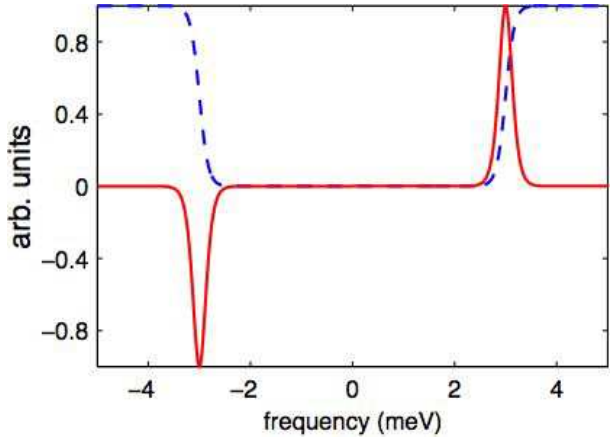}
\end{center}
\caption{(Color online) $\partial\delta N(0,\omega)/\partial\omega$ (solid) and $\delta N(0,\omega)$ (dashed) for $\omega_0=3$ meV and $T=1$ K ($m=\hbar=1$).}
\label{fig-dN}
\end{figure}
Next suppose that the position of the STM tip is kept fixed in space, i.e. letting $\bfr-\bfr_0=0$ where $\bfr_0$ is the position of the inelastic scattering center. Then, as show in the previous section, the derivative $\partial\delta N(0,\omega)/\partial\omega$ is expected to peak or dip at $\omega=\pm\omega_0$, c.f. Eq. (\ref{eq-ddN}) and (\ref{eq-dS}), as the frequency is varied. This is illustrated in Fig. \ref{fig-dN} (solid). These local extrema correspond to steps in $\delta N(0,\omega)$, which is also clear in Fig. \ref{fig-dN} (dashed).

\begin{figure}[t]
\begin{center}
\includegraphics[width=8cm]{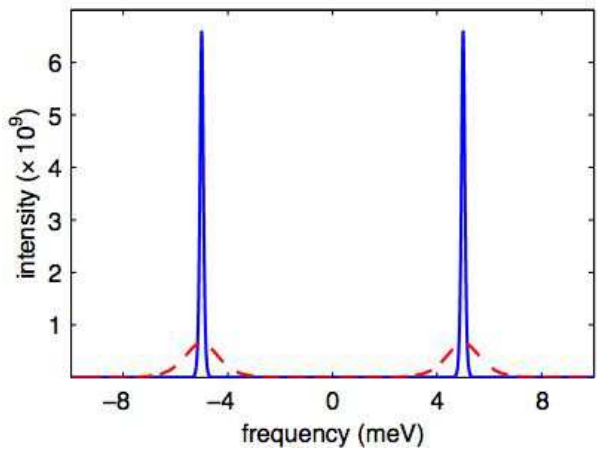}
\end{center}
\caption{(Color online) The derivative ($|\partial\delta N/\partial\omega|$)
 for different temperatures $T\in\{0.01,0.1\}$ K $\{$solid, dashed$\}$. The plots have been artificially broadened.}
\label{fig-dNT}
\end{figure}

From Eq. (\ref{eq-dS}) we find that intensity of the peaks in $|\partial\delta N/\partial\omega|$ is highly temperature dependent, which is illustrated in Fig. \ref{fig-dNT} for two temperatures. The plots clearly shows that the peaks are becomes sharper for decreasing temperatures, illustrating the $\delta$ function like behavior at low temperatures.

\section{Conclusions}
\label{sec-conclusions}
We propose a scanning technique that allows one to visualize
the oscillations in the inelastic scattering produced by the local
modes.  Standard way to image IETS of the molecule is to measure
$\partial^2I(\br, V)/\partial V^2$ locally near the scattering center.\cite{stipe1998,heinrich2004,grobis2005} We point out that the
"fingerprint" of the inelastic scattering will be present even away
from the scattering center. The incoming electron wave
donates/absorbs energy from the inelastic scattering center. In the
process of inelastic scattering the incoming and outgoing electron
waves interfere. This interference will can be clearly seen in the
real space oscillations of the second derivative $\partial^2I(\br, V)/\partial V^2$
with characteristic momentum $k_F$. The important point is that for
the flat density of states the interference between states that
undergone inelastic scattering will only be present at the bias $eV
= \hbar \omega_0$ and not at other energies. This makes the proposed
effect very different from the conventional Friedel oscillations
that occur for elastic scattering and are present for the wide range
of energies as the whole band of electrons participates in
screening. We  also argue that as long as the energy difference
between these states is small compared with the Fermi energy of
electrons, the dephasing in the process of scattering is small and
hence interference is preserved.

The technique, proposed  here,  can be applied to the inelastic
scattering of the electrons from the local vibrational modes of the
molecules on the surface \cite{grobis2005} and would offer exciting
possibilities to image the interference and interactions between
inelastic scattering processes in the ensemble of few scattering
centers. In practice the tunneling conductance in real materials will likely have peaks in $\partial^2I(\bfr,V)/\partial V^2$ as a function of position due to band structure. One would need to be able to separate these peaks from the peaks that arise due to inelastic scattering.

This work has been supported by US DOE, LDRD and BES, and was carried out under the auspices of the NNSA of the US DOE at LANL under Contract No. DE-AC52-06NA25396. We are grateful to M. Crommie,  J.C. Davis, D. Eigler, W. Harrison, H. Manoharan, and R. Wiesendanger for useful discussions.

\appendix
\section{Derivation of $\delta N$}
\label{app-dN} In the above analysis we neglected the real part of
the unperturbed Green's functions. However, as we show here, taking
this part into account does not significantly change the picture.
We have
\begin{widetext}
\begin{eqnarray*}
\delta N(\bfr,\omega)&=&
    -\frac{1}{\pi}\biggl(\frac{2\pi m}{\hbar^2}\biggr)^2J_0^2(k_Fr)\biggl\{
    \pi^2\im\Sigma(\omega)
    -\biggl[\im\Sigma(\omega)\log\left|\frac{\omega+D}{\omega-D}\right|
    -2\pi\re\Sigma(\omega)\biggr]
    \log\left|\frac{\omega+D}{\omega-D}\right|\biggr\}.
\end{eqnarray*}

Since we are interested in the frequency derivative of this expression we find that
\begin{eqnarray*}
\frac{\partial}{\partial\omega}\delta N(\bfr,\omega)&=&
    -\frac{1}{\pi}\biggl(\frac{2\pi m}{\hbar^2}\biggr)^2J_0^2(k_Fr)\biggl\{
    \pi^2\frac{\partial}{\partial\omega}\im\Sigma(\omega)
    -\biggl(\frac{\partial}{\partial\omega}\im\Sigma(\omega)\biggr)
    \log\left|\frac{\omega+D}{\omega-D}\right|
\\&&
    +\im\Sigma(\omega)\frac{4D}{\omega^2-D^2}
    \log\left|\frac{\omega+D}{\omega-D}\right|
    +2\pi\biggl(\frac{\partial}{\partial\omega}\re\Sigma(\omega)\biggr)
    \log\left|\frac{\omega+D}{\omega-D}\right|
    -2\pi\re\Sigma(\omega)\frac{2D}{\omega^2-D^2}
    \biggr\}
\end{eqnarray*}
Here,
\begin{eqnarray*}
\re\Sigma(\omega)&=&2\pi\lambda^2\frac{m}{\hbar^2}\int
    \biggl\{\frac{n_B(\omega_0)+f(\dote{})}{\omega+\omega_0-\dote{}}
        +\frac{n_B(\omega_0)+1-f(\dote{})}{\omega-\omega_0-\dote{}}\biggr\}d\dote{}
\\&=&
    2\pi\lambda^2\frac{m}{\hbar^2}\biggl\{
        n_B(\omega_0)
            \log\left|\frac{\omega+\omega_0+D}{\omega+\omega_0-D}\right|
        +[n_B(\omega_0)+1]
            \log\left|\frac{\omega-\omega_0+D}{\omega-\omega_0-D}\right|
        -2\omega_0\int\frac{f(\dote{})}{(\omega-\dote{})^2-\omega_0^2}d\dote{}
        \biggr\}
\end{eqnarray*}
which gives the derivative
\begin{eqnarray*}
\frac{\partial}{\partial\omega}\re\Sigma(\omega)&=&
    -2\pi\lambda^2\frac{m}{\hbar^2}\int
    \biggl\{\frac{n_B(\omega_0)+f(\dote{})}{(\omega+\omega_0-\dote{})^2}
        +\frac{n_B(\omega_0)+1-f(\dote{})}{(\omega-\omega_0-\dote{})^2}
            \biggr\}d\dote{}
\\&=&
    4\pi\lambda^2\frac{m}{\hbar^2}\biggl\{
        D\frac{n_B(\omega)}{(\omega+\omega_0)^2-D^2}
        +D\frac{n_B(\omega_0)+1}{(\omega-\omega_0)^2-D^2}
        +2\omega_0\int f(\dote{})\frac{\omega-\dote{}}
        {[(\omega-\dote{})^2-\omega_0^2]^2}
        d\dote{}\biggr\}
\end{eqnarray*}
\end{widetext}
However, since the logarithm $\log|(\omega\pm\omega_0+D)/(\omega\pm\omega_0-D)|\approx0$ for $D\gg|\omega\pm\omega_0|$ we see that all terms but the first are negligible in $\partial\delta N/\partial\omega$.

\end{document}